\documentclass[12pt,a4paper,leqno]{article}

\usepackage{cite}
\usepackage{amsmath,amssymb,amsfonts}
\usepackage{algorithmic}
\usepackage{graphicx}
\usepackage{textcomp}
\usepackage{setspace} 
\usepackage{lineno}
\usepackage{color,soul}
\usepackage{xcolor}
\def\BibTeX{{\rm B\kern-.05em{\sc i\kern-.025em b}\kern-.08em
    T\kern-.1667em\lower.7ex\hbox{E}\kern-.125emX}}

\usepackage{footnote}
\usepackage{textcomp}

\begin{document}
\begin{center}
\thispagestyle{empty}
\doublespacing
\large{ \textbf{
Importance of gravitational effects on the performances of a fully passive oscillating-foil turbine deployed horizontally}}
\vspace{10pt}
\normalsize

Alexina Roy-Saillant$^\text{a}$, Guy Dumas$^\text{a}$, Leandro Duarte$^\text{b}$, Guilhem Dellinger$^\text{b}$, Mathieu Olivier$^\text{a,}$\footnote{Corresponding author : mathieu.olivier@gmc.ulaval.ca} \\
$^\text{a}$ LMFN, Département de Génie mécanique, Université Laval,  Québec, Canada.\\
$^\text{b}$ Laboratoire ICube, Département de Mécanique, Strasbourg, France.

\end{center}

\newpage
\doublespacing
\begin{abstract}
\normalfont
This article presents a numerical study evaluating the impact of gravity on the performance of a fully passive oscillating-foil turbine operating in a horizontal configuration, that is, where the gravity acts along the heave direction. The study examines two sets of parameters corresponding to turbines driven by different aeroelastic instabilities. For turbines experiencing stall-flutter instability, the influence of gravity on performance metrics is minimal when inertial forces dominate or are comparable to gravitational forces. At high Froude numbers, buoyancy and weight are negligible, but their impact increases at lower Froude numbers, leading to reduced performance. Conversely,  turbines operating through coupled-flutter instability seem unsuitable for horizontal configurations since they require high foil moment of inertia and mass, which amplifies the effects of buoyancy and weight, thereby diminishing performance at any Froude number.
\end{abstract}

\vspace*{0.5em}

\newpage

\section{Introduction}
The current environmental context has heightened the importance of optimizing renewable energy extraction technologies.  While conventional hydropower remains a dominant source primarily through dam infrastructure, hydrokinetic turbines are gaining attention for their potential to reduce ecological disruption and adapt to diverse flow conditions.

As such, hydrokinetic turbines stand out for their versatility and ease of installation. Among them, numerous oscillating-foil concepts have emerged over the years \cite{Xiao2014, Young2014, Wu2020}. More specifically, passive oscillating-foil turbines have gained interest for their simplified mechanical design \cite{Peng2009, Zhu2009, Ma2021, veilleuxNumericalOptimizationFullypassive2017, boudreauParametric2020}. This concept is usually studied without considering gravitational effects on the turbine's dynamics. In such case, the optimal sets of parameters have been shown to yield efficiencies of 34\%~\cite{veilleuxNumericalOptimizationFullypassive2017} when the turbine undergoes stall-flutter instability and 53.8\%~\cite{boudreauParametric2020,guntherNumericalExperimentalComparison2022} when the turbine undergoes coupled-flutter instability. These studies revealed the potential of energy-extracting mechanisms taking advantage of coupled flutter, a concept further supported by subsequent studies~\cite{goyaniukEnergyExtractionPotential2023}. While these results remain valid for horizontal deployments under the assumption of neutral buoyancy~\cite{zhangDualfunctionFlappingHydrofoil2024}, this assumption becomes limiting when foil mass is treated as a design variable. In such cases, neglecting gravitational effects may prevent the identification of truly optimal parameter sets, thereby constraining the system’s performance potential.

\begin{figure}[htb!]
    \centering
    \includegraphics[width=0.99\textwidth]{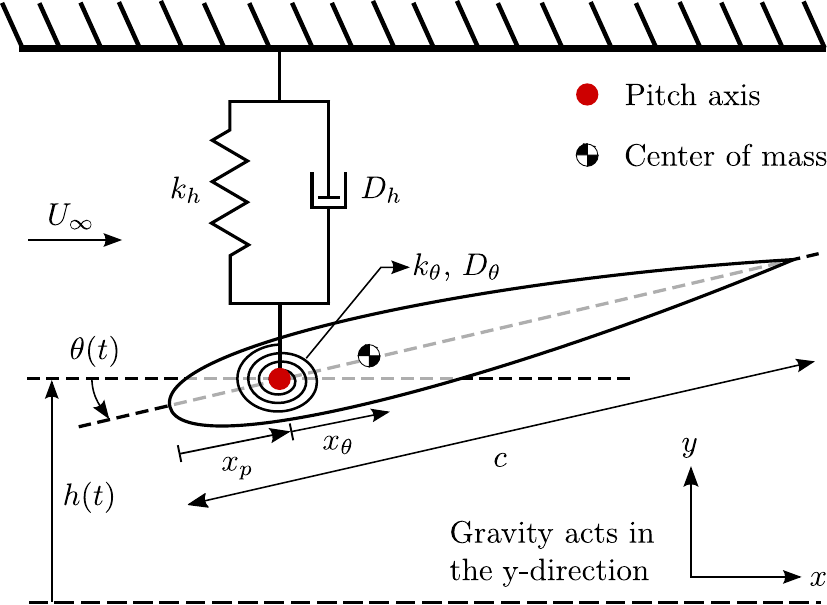}
    \caption{The fully passive oscillating-foil turbine prototype, adapted from~\cite{boudreauOptimizing2019}. } 
    \label{fig:schema_HAO}
\end{figure}

As shown in Fig.~\ref{fig:schema_HAO}, the turbine consists of a NACA~0015 foil mounted on springs at its pivot center, allowing for two degrees of freedom: pitching and heaving. Its motion is driven by the interactions between the flow and the elastic supports. The electric generator acts as a damper in both degrees of freedom. In this horizontal configuration, gravity affects the non-linear dynamics of the turbine. This paper thus proposes an approach to study the concept's viability in the horizontal configuration by considering the gravitational force in the mathematical model that takes into account fluid-structure interactions. As turbine performance is known to be sensitive to confinement~\cite{guntherNumericalExperimentalComparison2022,mannConfinementEffectsHydrodynamic2022}
, the proximity of the riverbed is also considered as it has been shown to affect performance, although it will not be discussed in detail in this paper.

\section{Dynamics and metrics}
Since the Fully Passive Oscillating-Foil Turbine (FP-OFT) is studied in its horizontal configuration, the forces and moment equations must include the gravitational terms. To reproduce previously found efficient vertical configurations~\cite{veilleuxNumericalOptimizationFullypassive2017,boudreauParametric2020}, pre-stretches need to be added to the springs to ensure equilibrium around $h=0$ and  $\theta=0$ in a stagnant flow. The system of equations is then
\begin{equation}
\label{eq:f_et_m_avec_ini}
\begin{aligned}
      F_{y}&= m_{h} \ddot{h} +S(\ddot{\theta} \cos \theta - \dot{\theta}^2 \sin \theta) + D_h \dot{h}~+~k_{h} (h + h_{0}) + m_{h} g,\\
      M &= I_{\theta} \ddot{\theta} + S(\ddot{h} + g) \cos \theta + D_{\theta} \dot{\theta} + k_{\theta} (\theta + \theta_{0}),
\end{aligned}
\end{equation}
where $h_0$ and $\theta_0$ correspond to the pre-stretches. The mass imbalance is defined as $S = m_{\theta} x_{\theta}$. Adding pre-stretches to the springs will lessen the effect of gravity but will not be sufficient to counteract the asymmetrical nature of the problem. While the foil rotates, the horizontal distance between the center of mass and the geometric centroid, where the buoyancy force and the weight are respectively applied, varies. Since the pre-stretches cannot change over time, it is not possible to completely counteract these forces over the oscillation period.

The problem is governed by numerous parameters. The parameters related to the flow are the inlet velocity $\mathit{U_{\infty}}$, the fluid density $\rho$, and the fluid kinematic viscosity $\nu$. The properties of the installation site such as the distance between the foil and the riverbed $\mathit{L_{b}}$ and gravity $\mathit{g}$ also impact the behavior. The pertinent structural parameters are the chord length $\mathit{c}$, the mass in translation $\mathit{m_h}$, the damping factors in both directions $\mathit{D_h}$ and $\mathit{D_\theta}$, the spring coefficients $\mathit{k_h}$ and $\mathit{k_\theta}$, the system inertia $I_{\theta}$, and the mass imbalance $\mathit{S}$~\cite{boudreauParametric2020}. Since the study is conducted in 2D, the extensive properties are defined here as their value per unit depth. The dimensional analysis of the problem yields the following 10 dimensionless numbers, the first ones being the Reynolds and the Froude numbers:
  \begin{equation}
  Re=\frac{U_{\infty} c}{\nu},
  \end{equation} 
 \begin{equation}
  Fr_{c}=\frac{U_{\infty}}{\sqrt{gc}}.
	\label{Frcorde}
  \end{equation} 
All calculations are done at $Re=3.9 \times 10^6$, ensuring fully turbulent boundary layers~\cite{boudreauParametric2020,guntherEfficientControlFully2024}. The parameters related to the structure~\cite{boudreauParametric2020}, along with the distance to the riverbed $L_b$, must also be considered:
\begin{equation*}
\begin{aligned}
    m_h^* &=\frac{m_h}{\rho c^2},\quad
    D_h^*=\frac{D_h}{U_{\infty}\rho c},\quad
    k_h^*=\frac{k_h}{U_{\infty}^2\rho},\quad
    L_{b}^*=\frac{L_{b}}{c}, \\
    I_{\theta}^* &=\frac{I_{\theta}}{\rho c^4}, \quad
    D_{\theta}^*=\frac{D_{\theta}}{U_{\infty}\rho c^3},\quad
    k_{\theta}^*=\frac{k_{\theta}}{U_{\infty}^2\rho c^2},\quad
    S^*=\frac{S}{\rho c^3}.\quad
\end{aligned}
\end{equation*}

Efficiency is used to quantify the performance of the FP-OFT. It can be obtained with the power coefficient at the generator $C_{P}$. Considering a linear generator acting as a linear damper in both the heaving and pitching equations of motion, the power coefficient yields~\cite{boudreauParametric2020}
\begin{equation}
\begin{aligned}
    \overline{C_{P}}&= \overline{C_{P_{D_{h}}}} + \overline{C_{P_{\theta}}}\\ 
                &= \int_{t_i}^{t_i+T} \frac{1}{T} \frac{D_{h} \dot{h}^2}{0.5 \rho U_{\infty}^3 c} \,dt 
                + \int_{t_i}^{t_i+T} \frac{1}{T} \frac{D_{\theta} \dot{\theta}^2}{0.5 \rho U_{\infty}^3 c} \,dt.
\end{aligned}
\end{equation}
The turbine efficiency is then defined as
\begin{equation}
\label{eq:efficacité}
\eta = \frac{\overline{C_{P}} \cdot c}{d},
\end{equation}
where $d$ is the total distance swept by foil over a complete cycle and $T$, the oscillation period. This result is averaged over at least the last 10 cycles when a permanent regime is reached. 

The response of the system can also be characterized by the ratio of the different forces at play, approximated by the order of magnitude of each relevant force. The inertial forces $F_I$ scale with the heaving motion, hence by the translating mass and its acceleration, estimated by the frequency $\omega$ and the heaving amplitude $H_0$ ($\ddot{h} \approx \omega^2 H_0$). The pressure forces $F_P$ scale with the dynamic pressure associated to the heaving motion ($\rho (\omega H_0)^2$). The gravity force $F_g$ scales with the weight of the translating mass. The spring force $F_s$ scales with the recalling force generated when the foil is at heaving oscillation amplitude. The buoyancy force $F_B$ scales with the weight of the displaced fluid volume $V$. This yields the following ratios:
\begin{equation*}
\begin{aligned}
    \frac{F_P}{F_I} &=\frac{\rho H_0 c }{m_h}, \qquad
    \frac{F_{I}}{F_{g}} =\frac{\omega^2 H_0}{g}, \qquad
    \frac{F_{s}}{F_{P}} =\frac{k_h}{\rho H_0 \omega^2 c}, \\
    \frac{F_{B}}{F_{s}} &=\frac{\rho g V}{k_h H_0}, \qquad
    \frac{F_{g}}{F_{s}} =\frac{m_h g}{k_h H_0}, \qquad
    \frac{F_{B}}{F_{g}} =\frac{\rho V}{m_h}.
\end{aligned}
\end{equation*}
These dimensionless ratios will thus be useful for interpretations in the analysis further below.

\section{Numerical methodology}
\subsection{Models and methods}
A 2D CFD computing environment was devised to simulate the turbine under different conditions, including the Froude number of the flow (\ref{Frcorde}), which is used to quantify the effect of gravity on the turbine dynamics. These simulations are conducted within Star-CCM+, and they use an in-house fluid-structure interaction algorithm based on the one used in previous studies involving the vertical configuration of the FP-OFT~\cite{veilleuxNumericalOptimizationFullypassive2017, boudreauParametric2020, guntherEfficientControlFully2024, olivierStrongFluidSolid2019}.
This coupling algorithm is based on Broyden's method, a quasi-Newton method generally used to solve non-linear systems. The method is applied on the equations of motion (\ref{eq:f_et_m_avec_ini}) in which the external force and moment generated by the flow are obtained through the CFD solver. This implicit method remains stable even when the added-mass effect is significant, as in the studied case.

The CFD solver is configured to solve the incompressible Navier-Stokes equations. To account for the turbulent nature of the flow at the Reynolds number considered, an unsteady Reynolds averaged (URANS) approach is used along with the $k-\omega$ SST turbulence model. The turbulence model is used without wall functions. As such, the mesh resolution at the wall is adjusted to capture the viscous sublayer of the boundary layers.
Lastly, gravitational effects are taken into account simply by adding the gravitational force in both the equations of motion and the Navier-Stokes equations.

\subsection{Boundary conditions}
The piezometric pressure is set to 0 at the outlet and the velocity of the fluid is set as $U_{\infty}$ at the inlet. To reproduce clean flow conditions at the inlet, the turbulent intensity is set to a low value of $I=0.01$ and the turbulent viscosity ratio is set to $\mu_t/\mu \approx 0.21$ to correspond to $\tilde{\nu}/\nu = 3$~\cite{vogelSpalartAllmarasTurbulenceModel2023, spalartEffectiveInflowConditions2007}. More details on the boundary conditions are provided in Fig.~\ref{fig:mesh}. 

\subsection{Mesh description and verification}
\begin{figure}[htb!]
    \centering
    \includegraphics[width=0.99\textwidth]{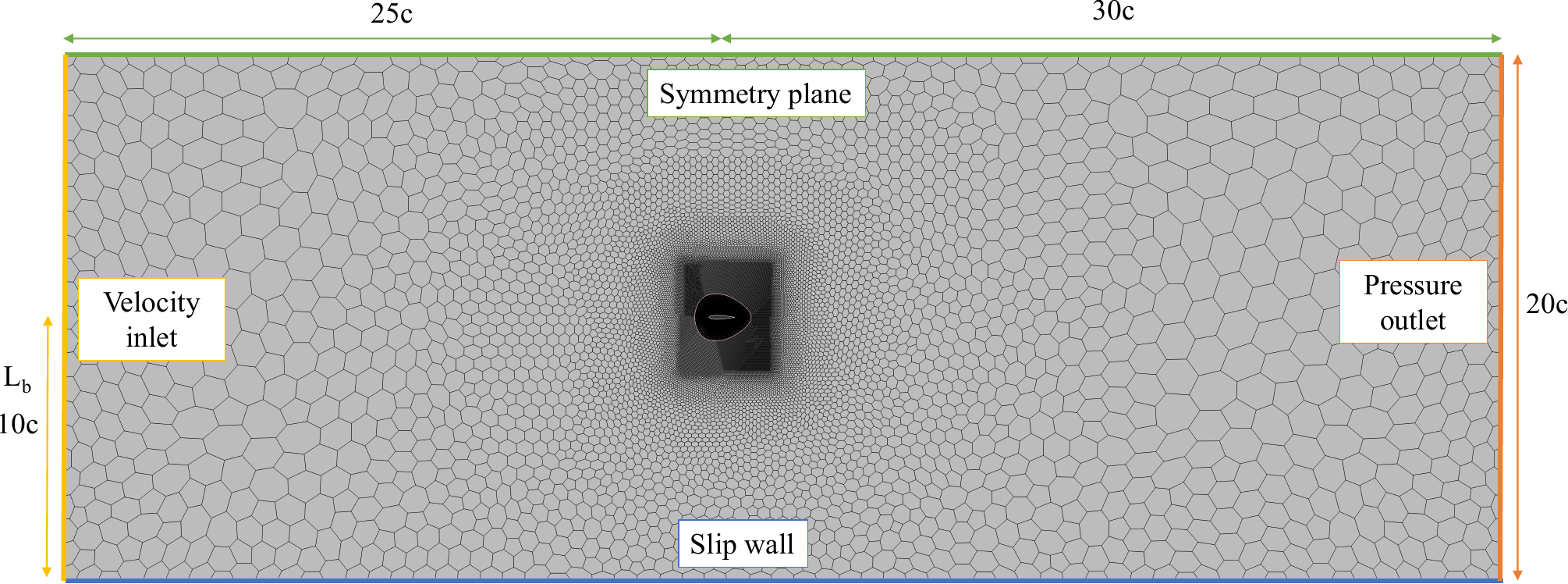}
    \caption{Boundary conditions.}
    \label{fig:mesh}
\end{figure}

The mesh dimensions, shown in Fig.~\ref{fig:mesh}, were chosen so that the confinement of the domain is representative of that of a typical river~\cite{niPerformanceHydrofoilOperating2021,gauvintremblayEffetsConfinementSurface2015, esmaeilifarHydrodynamicSimulationOscillating2017}. The present study is conducted at a realistic confinement level for a river: the distance between the riverbed and the foil at its original position is fixed at $L_{b} = 10c$, and the mesh is built so that the foil is vertically centered.

The cell size was chosen to ensure that the results are independent of the grid size. Since the foil is moving, the cells around it are part of a moving mesh that is overlaid on top of a background mesh. The mesh in the overset region is constructed to ensure near wall cells lie in the viscous sublayer ($y^+ \sim 1$). Cells in the background mesh are refined in the predicted foil moving zone to avoid resolution discrepancies at the interface with the overset region.

Different meshes were tested to assess convergence. A typical simulation was studied with overset meshes of 250, 368, 500, and 750 nodes around the foil and the corresponding background meshes. Information related to the different meshes is compiled in Table~\ref{tab:convergence_grille}.

\begin{table}[htb!]
	\caption{Mesh characteristics.}
	\begin{center}
		\begin{tabular}{ccccc}
			\hline
			Nb. of nodes & Base size & Refined cell size & Nb. of cells & $\Delta t U_{\infty}/c$    \\
			\hline
			250             & 0.4            & 0.04                          & 166462             & 0.015 \\
368             & 0.35           & 0.035                         & 397609             & 0.01  \\
500             & 0.3            & 0.03                          & 476647             & 0.008 \\
750             & 0.15           & 0.0225                        & 1113036            & 0.005 \\
		\hline
		\end{tabular}
		\label{tab:convergence_grille}
	\end{center}
\end{table}

\begin{figure}[htb!]
    \centering
    \includegraphics[width=0.99\textwidth]{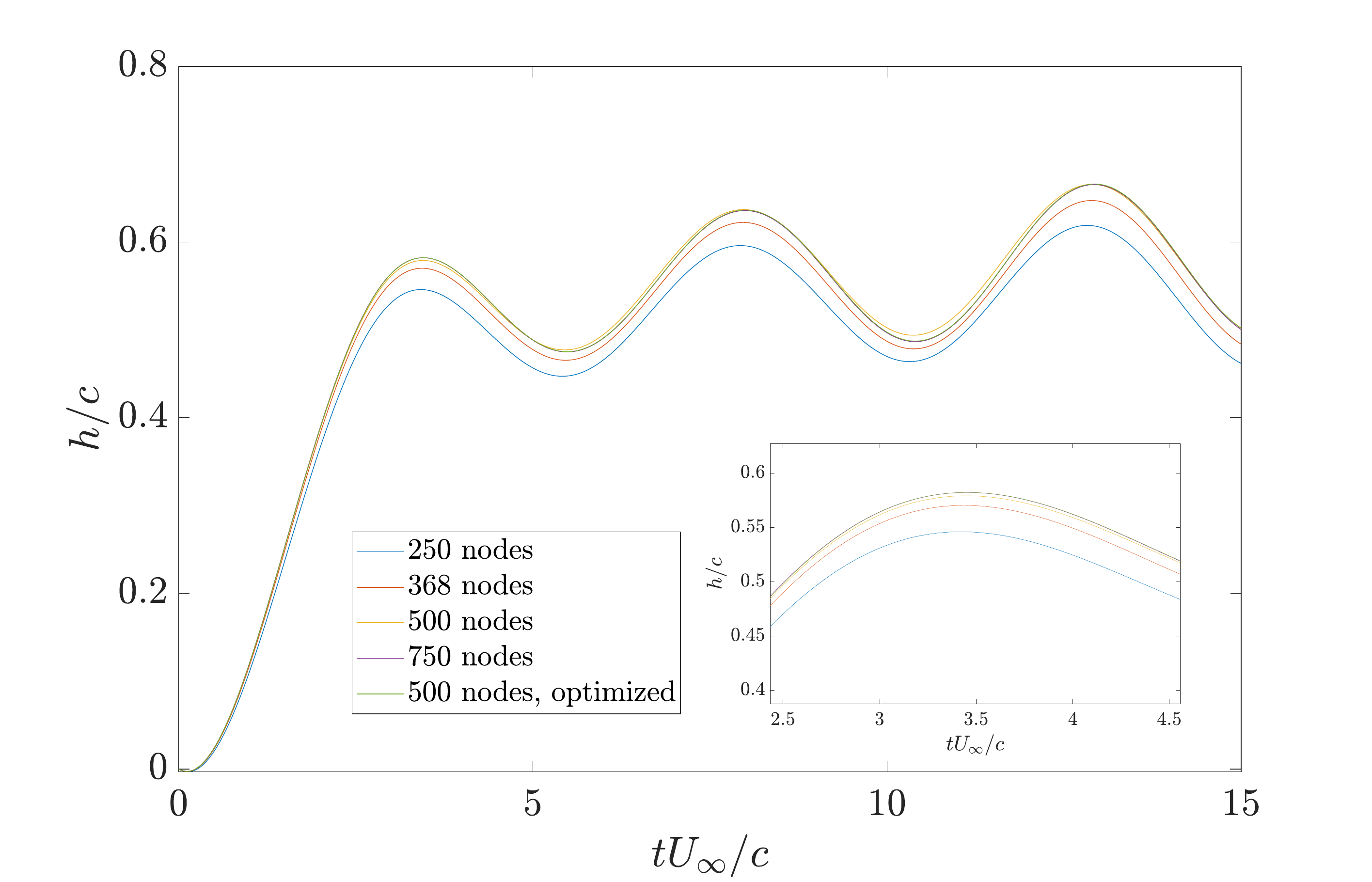}
    \caption{Position as a function of time for all tested meshes.}
    \label{fig:y_n_opt}
\end{figure}

Fig.~\ref{fig:y_n_opt} shows that the 500-node mesh provides sufficient precision for the proposed study. From there, the background grid was further optimized to account for the foil motion (which was not known prior to running the simulations). This final mesh and corresponding time step were used throughout the study.

\subsection{Validation}
Since the FP-OFT is essentially a forced mass-spring-damper system, the analytical solution of its movement can be easily determined if the system is considered in a theoretical vacuum (without fluid forces) and if the system is decoupled ($S^*=0$). Indeed, in this case, both the heaving and pitching motions follow a simple harmonic motion.
This solution was then compared with the one obtained by the algorithm.
The numerical results agree with the analytical solution for both the vertical and angular positions with a maximum absolute difference over time between the numerical and analytical solutions of less than 0.2\% of the corresponding maximum amplitude.

\begin{table}[htb!]
\centering
\caption{Dimensionless parameters used in the stall-flutter case~\cite{veilleuxNumericalOptimizationFullypassive2017}.}
\label{tab:casoptiVeilleux_validation}
\begin{tabular}{cc}
\hline
Parameter      & Value \\
\hline
$m_h^*$        & 3.036  \\
$k_{\theta}^*$ & 0.031  \\
$D_{\theta}^*$ & 0.119  \\
$I_{\theta}^*$ & 0.095  \\
$S^*$          & -0.029 \\
$k_{h}^*$      & 1.206  \\
$D_{h}^*$      & 1.501  \\ 
$\frac{x_{p}}{c}$      & 0.33  \\
\hline
\end{tabular}

\end{table}

To further validate the implementation, the results were also compared successfully with the optimal case obtained in \cite{veilleuxNumericalOptimizationFullypassive2017}. This case is driven by the stall-flutter instability and served as a baseline case in this study. The corresponding parameters are listed in Table~\ref{tab:casoptiVeilleux_validation}.
Moreover, previous comparisons between experimental and CFD studies of the FP-OFT~\cite{guntherNumericalExperimentalComparison2022} have shown that 2D and 3D URANS simulations can achieve performance predictions similar to the experimental ones. While two-dimensional studies introduce additional modeling error, they still provide proper trends of the performance metrics when compared with 3D moderate- to high-aspect-ratio cases.

\section{Results}
\subsection{FP-OFT operating via stall-flutter instability}
To quantify the impact of gravity on the performance of the FP-OFT, the turbine is tested at different $Fr_c$. This study can then be used as a guide to select the chord length depending on the implantation site, as the gravitational constant $g$ cannot be changed, and the velocity of the flow is generally fixed at a designated site.

\begin{figure}[htbp!]
    \centering
    {\includegraphics[width=0.99\textwidth]{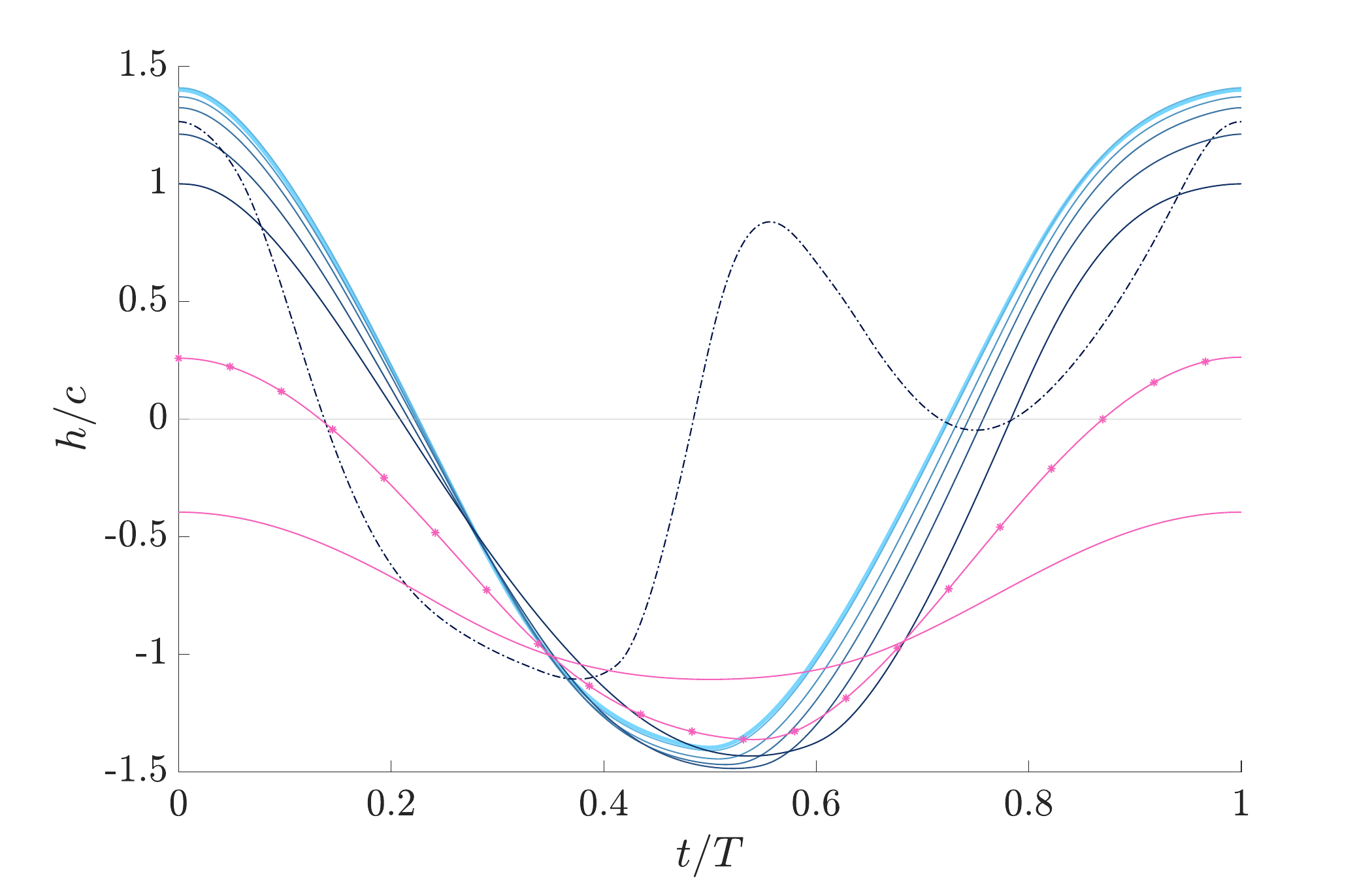}}
    \\
    {\includegraphics[width=0.99\textwidth]{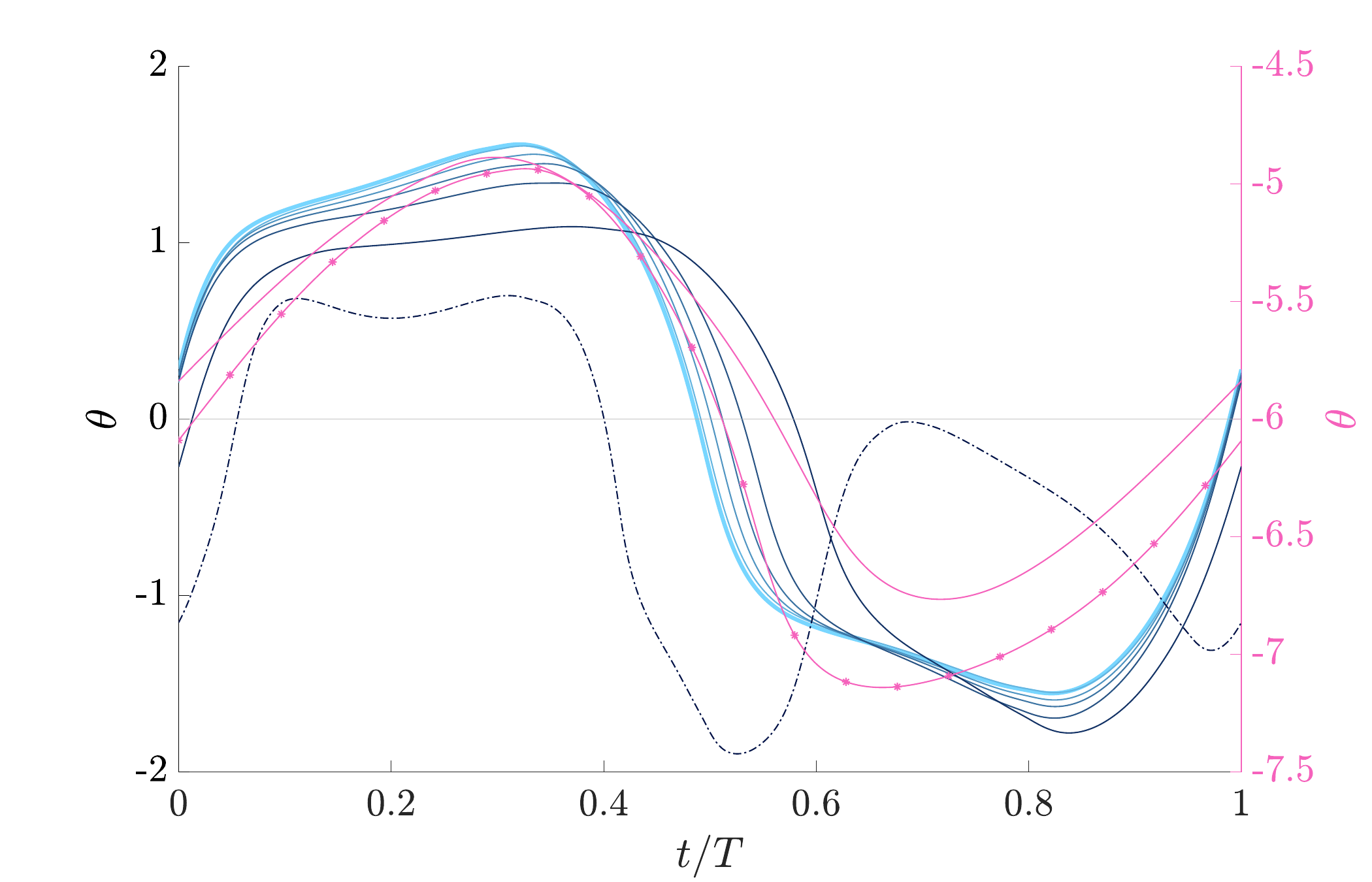}}
    \\
    {\includegraphics[width=0.99\textwidth]{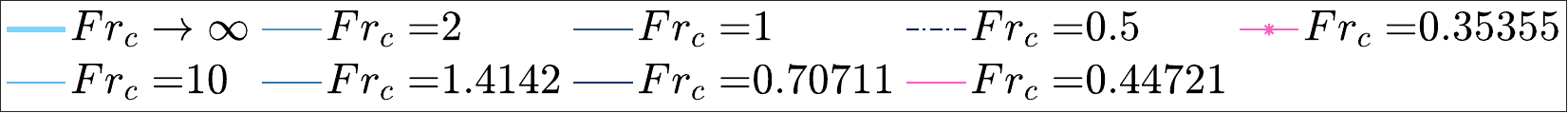}}
    \caption{Vertical position (top) and angular position (bottom) of the foil over a cycle for different Froude numbers. Pink curves correspond to cases where the foil oscillates around $\theta \geq 2\pi$.}
    \label{fig:varGravite_stall}
\end{figure}

Fig.~\ref{fig:varGravite_stall} shows the normalized response of the foil over a cycle at different Froude numbers. One can easily observe that gravity impacts the foil's movement, which in turn affects the FP-OFT's performance.

The impact of gravity on the system is related to the ratio between the moment generated by the buoyancy force and the mass imbalance.
Firstly, since the mass imbalance is negative, the center of mass is located upstream of the pivot center (see Fig.~\ref{fig:schema_HAO}), and the weight of the foil generates a counter-clockwise rotation around the pitch axis. The buoyancy force, which is always applied at the geometrical center of the foil (located at $x/c=0.4204$ for a NACA~0015), also produces a counter-clockwise rotation. Thus, the forces generated by the presence of gravity produce a counter-clockwise moment when the foil is horizontal. The pre-stretch applied to the pitching spring produces a clockwise moment of equal value. As stated previously, the pre-stretches do not vary over the cycle, while the moment generated by the buoyancy force and the static imbalance do. This phenomenon, along with the asymmetrical nature of the problem, explains the asymmetrical movement of the foil over a cycle seen in Fig.~\ref{fig:varGravite_stall}: the sudden angular acceleration at the end of the ascent of the foil and the slowdown of the angular movement of the descending foil.

When gravitational effects become significant compared to inertial effects, i.e., when $Fr_{c} \rightarrow 0$, the influence of pre-stretches becomes increasingly important. The foil is more stable at positive angles than at negative angles: the descent of the foil, during which it rotates counter-clockwise, lasts longer than its ascent. When the foil is vertical, i.e., when $\theta \simeq \pi/2$, the buoyancy force and the weight of the foil do not generate any moment on the pitching axis,  but the spring pre-stretches are still compensating for the moments that are no longer acting. It is thus at these points in the cycle (during ascent and descent) that the pre-stretch of the pitch spring is no longer suitable and has a more marked effect, resulting in moderate angles being reached more slowly as the foil descends, and higher angles being reached more abruptly as it rises. The effects on angular position are also reflected in vertical position: the foil descends more slowly than it rises. Gravity increases the relative magnitude of the mass imbalance, weight, and buoyancy to the other forces and moments on the foil, and thus the pre-stretching required to counteract them. This explains the intensification of the effects observable when the Froude number decreases, i.e., when the magnitude of gravitational effects increases relative to that of inertial effects.

\begin{figure}[htb!]
    \centering
    \includegraphics[width=0.99\textwidth]{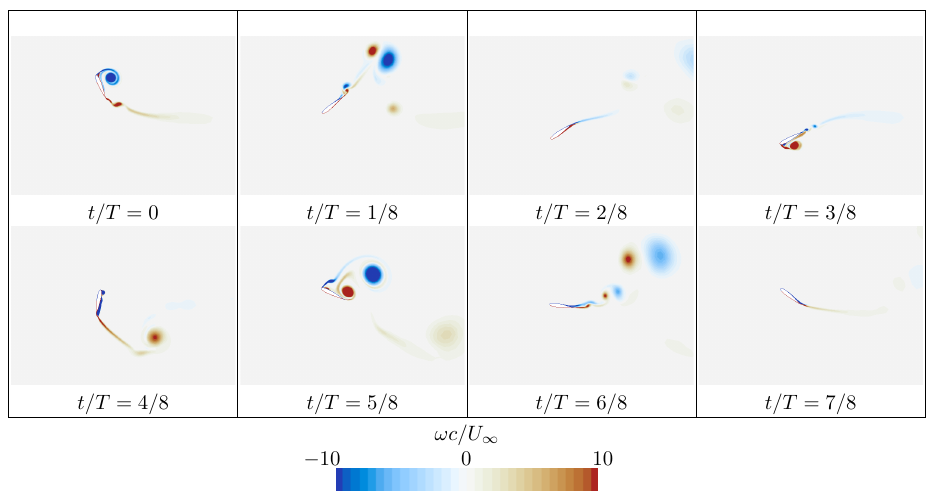}
    \caption{Vorticity around the foil if $Fr_c = 0.5$.} 
    \label{fig:vort_g4_stall}
\end{figure}

The case with $Fr_c=0.5$ is characterized by a marked change in the cycle. Whereas at higher Froude numbers the oscillation was quasi-sinusoidal, at $Fr_c=0.5$ the oscillation takes place in two stages. Halfway through the cycle ($t/T=1/2$, see Fig.~\ref{fig:vort_g4_stall}), the foil rotates slightly above $\theta=\pi/2$. Since the center of mass is located upstream of the pivot point ($S$ is negative), if the foil overshoots the vertical axis, the moment generated by its weight and buoyancy becomes clockwise, accelerating its rotation to complete a full revolution. At $Fr=0.5$, the overshoot is very slight and can be countered by the springs. A two-stage foil oscillation is then observed, caused by the foil stalling in the middle of the cycle, as it rises and slightly overshoots the vertical, followed by a secondary stall at the end of the course. 

However, at $Fr_c < 0.5$, the springs can no longer counteract the weight and buoyancy forces as well as the moments these forces generate when the foil reaches an angle greater than $\pi/2$ during its ascent. The foil then makes a complete turn on itself during its first ascent, and oscillates around a new equilibrium position at an angle greater than $2\pi$ afterwards. Since the torsion spring operates at $\theta \geq 2\pi$, the force it generates is greater on the cycle and the motion is damped.

\begin{figure}[htb!]
    \centering
    \includegraphics[width=0.99\textwidth]{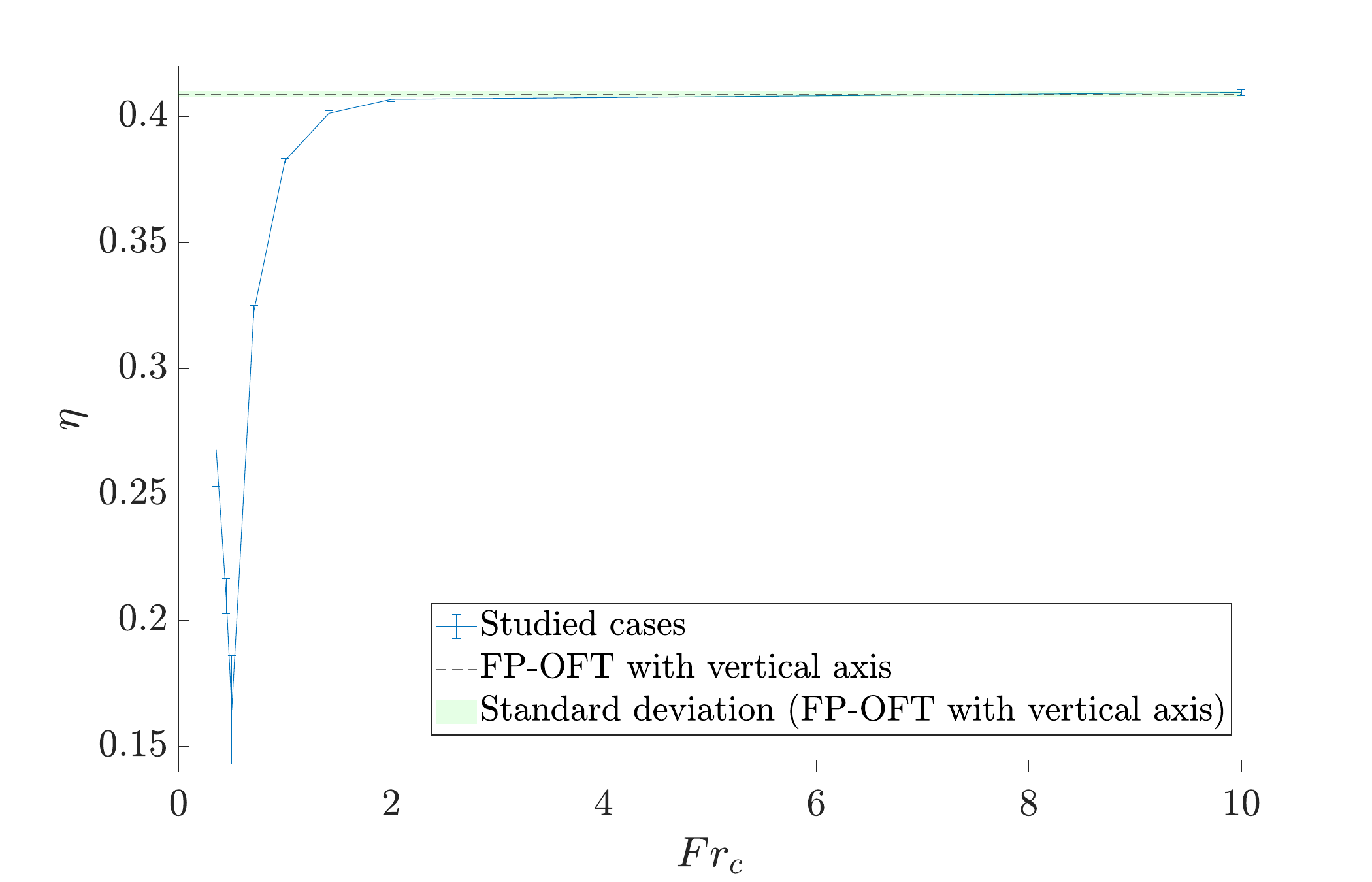}
    \caption{Efficiency of the FP-OFT at different $Fr_{c}$ } 
    \label{fig:rendement_Fr_c_stall}
\end{figure}

The effects of the Froude number on the trajectory also indicate a change in the performance of the FP-OFT. It is generally observed that, as gravitational effects become more important than inertial effects, i.e., as the Froude number $Fr_{c}$ decreases, the efficiency shown in Fig.~\ref{fig:rendement_Fr_c_stall} decreases. However, at $Fr_{c} < 0.5$, this trend reverses. Although these Froude numbers result in slightly higher efficiencies, these cases are not investigated.
Indeed, in typical river conditions, the length of the foil's chord $c$ would have to be very long, on the order of 10 meters, for $Fr_c$ to be that small, and thus these cases will most likely not be encountered. 

\begin{figure}[htb!]
    \centering
    {\includegraphics[width=0.99\textwidth]{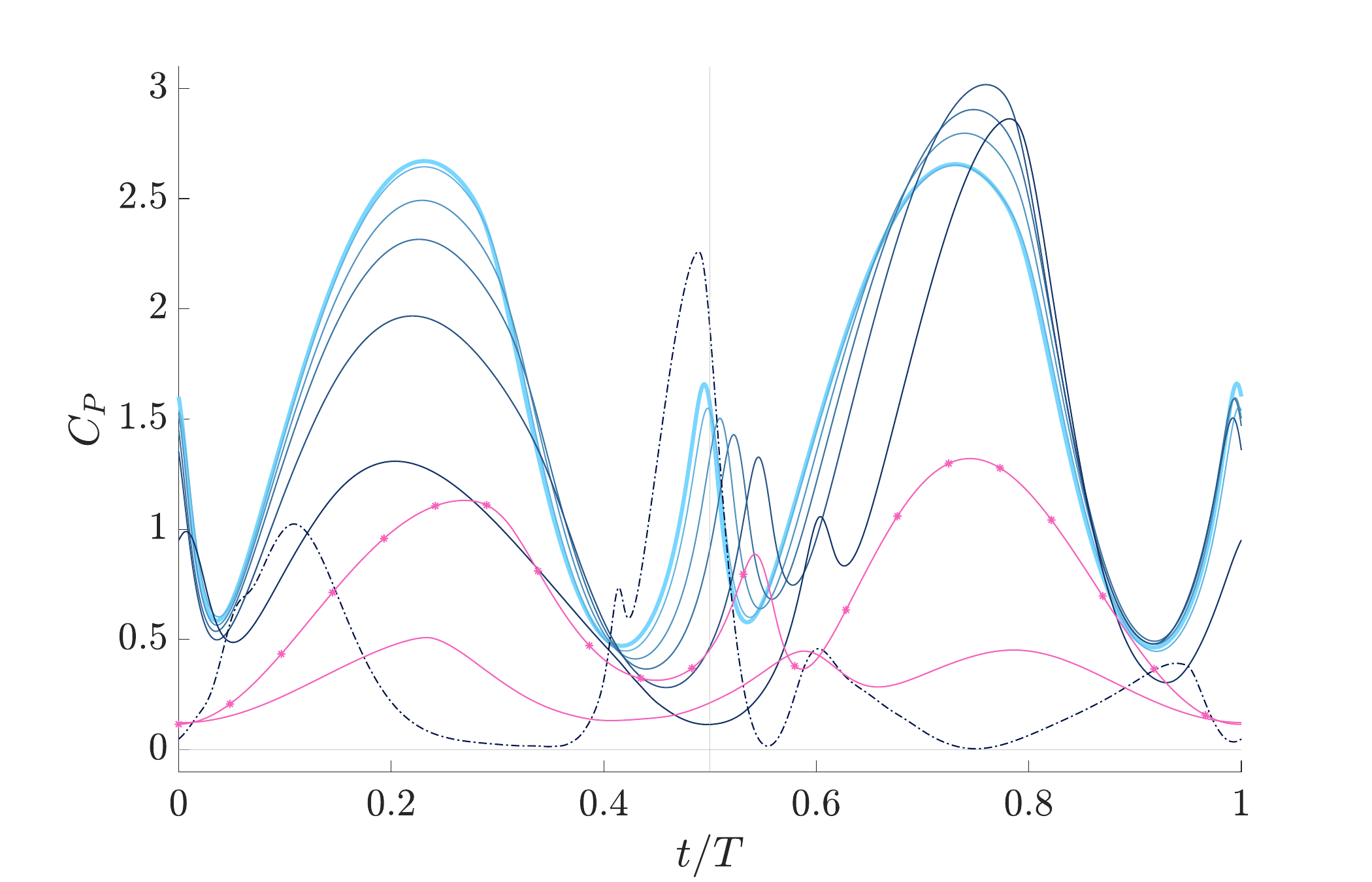}}
    \\
    {\includegraphics[width=0.7\textwidth]{legende_stall_g.pdf}}
    \caption{Value of $C_{P}$ over a cycle.}
    \label{fig:Cp_gravite_stall}
\end{figure}

The study of $C_{P}$ over one cycle (Fig.~\ref{fig:Cp_gravite_stall}) shows the evolution of the power coefficient over the period. For $Fr_c > 0.5$, the power decreases as the foil descends, compared to the case in a vertical configuration, but increases as it rises. These changes are of the same order of magnitude, explaining the small change in efficiency. However, for $Fr_c < 1.0$, the cycle becomes more asymmetric, and the abrupt descent of the angle of attack, or even the two-stage oscillation for the case with $Fr_c = 0.5$, creates a significant loss of power and thus of efficiency. The cases with $Fr_c < 0.5$, as shown in Fig.~\ref{fig:Cp_gravite_stall}, provide much lower power, but begin to recover the efficiency of $Fr_c > 0.5$ cases when the $Fr_c$ becomes very small and the foil oscillates around a new equilibrium point greater than $\theta = 2\pi$.

At all $Fr_c$ studied, the forces of pressure and inertia are of similar orders: both are therefore involved in foil motion, and greatly dominate over the force of the heave spring and, in the majority of cases, gravitational effects. If $Fr_c$ decreases, gravitational effects gain in importance over inertial effects. The spring force gradually becomes of the same order as the buoyancy force, both of which are then less important than gravitational effects. This explains the abrupt drop in efficiency of the oscillating foil at around $Fr_c < 1$. It can be determined that the weight is dominant over the buoyancy force in this case, for all $Fr_c$: $F_B/F_g=0.0338$. The other ratios are compiled in Table~\ref{tab:Pt_stall_g}. 

\begin{table}[htb!]
\centering
\caption{Force ratio at different $Fr_{c}$.}
\label{tab:Pt_stall_g}
\begin{tabular}{cccccc}
\hline
$Fr_{c}$ & $F_{I}/F_{g}$ & $F_{P}/F_{I}$ & $F_{s}/F_{P}$ & $F_{B}/F_{s}$ & $F_{g}/F_{s}$ \\
\hline
0.5000   & 23.8233    & 0.2675 & 0.0127 & 0.4198 & 12.4032 \\
1.0000   & 131.7072   & 0.4472 & 0.0092 & 0.0628 & 1.8549  \\
2.0000   & 521.3016   & 0.4641 & 0.0093 & 0.0151 & 0.4468  \\
$\rightarrow \infty$ & $\rightarrow \infty$   & 0.4638 & 0.0093 & 0.0000 & 0.0000  \\
\hline
\end{tabular}
\end{table}

The relative importance of the forces can thus be summarized as follows:
\begin{itemize}
\item[-] For $Fr_c \gtrsim 1$:
\begin{equation}
\label{res_Fr_grt_1}
F_B < F_g < F_{s} < F_P \sim F_I.
\end{equation}
\item[-] For $Fr_c \lesssim 1$:
\begin{equation}
\label{res_Fr_less_1}
F_{s} \sim F_B < F_g \sim F_P \sim F_I.
\end{equation}
\end{itemize}

\subsection{FP-OFT operating via coupled-flutter instability}
Some studies have introduced different parameters for the FP-OFT, enabling it to undergo coupled-flutter instability. In a vertical configuration, these turbines can achieve higher efficiencies compared to those experiencing stall-flutter instability~\cite{boudreauParametric2020}.
However, it can be determined from the outset that efficiency will decrease for these cases when the turbine operates in a horizontal configuration with the presence of gravity. Indeed, the parameters associated with coupled-flutter instability typically include very high foil inertia and mass, making the buoyancy force and weight non-negligible in the system response.
Hence, since the effects of $F_B$ and $F_g$ gain importance, gravitational effects are expected to decrease the turbine performances (see \ref{res_Fr_grt_1} and~\ref{res_Fr_less_1}).
This section confirms this trend and further highlights why.

To ensure that the buoyancy force and the weight cancel each other throughout the cycle, the density of the foil must be uniform and of the same value as the fluid in which the foil is placed. This can be expressed as:
\begin{equation}
\label{eq:Ksi}
\frac{m_{\theta}}{\rho V} = 1.
\end{equation}
In the case of coupled-flutter instability, the position of the pivot center is $x_p/c=0.25$, whereas the position of the center of mass of the foil, for the NACA~0015 profile, is $x_{cm}/c=0.4204$.
Therefore, the distance between the center of mass of the foil and the pivot center, as shown in Fig.~\ref{fig:schema_HAO}, is  $x_{\theta}/c=(x_{cm}-x_p)/c=0.1704$,
and if $\tilde{I}$ is the inertia of the NACA~0015 profile, the dimensionless moment of inertia of the foil is
\begin{equation}
I_{\theta}^* =\frac{m_{\theta}}{\rho c^2} \left( \frac{\tilde{I}}{c^2} + \frac{x_{\theta}^2}{c^2}\right) = 0.003039.
\end{equation}

This puts a constraint on the following parameters to ensure that the gravity has very little impact on the performances:
\begin{equation}
S^*= 0.01751,\qquad
m_{\theta}^*= 0.1027.
\end{equation}
Thus, the conditions that must be met to cancel out the effects of gravity imply a constraint on the value of the foil's moment of inertia $I_{\theta}$. These conditions also limit the value of the mass imbalance $S$, since the mass of the foil $m_{\theta}$ must have a certain value, and the position of the pivot point is fixed. The value of the corresponding mass imbalance is significantly lower than that of the efficient cases undergoing coupled-flutter instability in previous studies~\cite{boudreauParametric2020}. The static moment $S$ links the equations of forces and moments on the foil (see \eqref{eq:f_et_m_avec_ini}). Since $S$ is very small, the system is almost decoupled and the pitching and heaving motions are mostly independent. However, the coupled-flutter instability, as its name suggests, relies on the system being coupled and the motions being interdependent. This predicts reduced performance of the FP-OFT operating with coupled-flutter instability in a horizontal configuration with gravity, as reported in \cite{roy-saillantEtudeLimpactSurface2024}. Conversely, FP-OFTs in a horizontal configuration experiencing stall-flutter instability perform better due to the nature of the instability governing their motion and the structural parameters involved. Their moment of inertia and mass imbalance can be low, as this instability relies less on the coupling of dynamic equations compared to coupled-flutter instability. 

\section{Conclusion}
This paper demonstrated the importance of gravitational effects in the dynamics of the FP-OFT operating in a horizontal configuration despite the addition of pre-stretches in the springs. Indeed, buoyancy and weight are purely vertical forces applied to the foil's geometric center and center of mass, respectively. As the foil rotates, the moment generated by these forces changes, and the pre-stretch calculated at a zero angle is no longer sufficient to cancel out the gravitational effects.

The study confirms that a performant stall-flutter-driven case can be viable even in a horizontal configuration, taking gravitational effects into account. There is very little reduction in performance as long as the Froude number remains high enough, i.e., as long as the inertial effects are of the same order as the gravitational effects ($Fr_c \gtrsim 1$), which can be controlled by judicious choice of chord length with respect to the flow velocity at the installation site. This ensures that the buoyancy force and the weight of the foil are less significant than the other forces at play in the foil's dynamic response. The velocities observed in rivers, together with terrestrial gravity, make it easy to achieve $Fr_c \gtrsim 1$, at which the performance of the FP-OFT is only marginally affected. It is therefore reasonable to be optimistic about the viability of the turbine operating with stall-flutter instability in a horizontal configuration under typical river conditions.

On the other hand, the parameters corresponding to coupled-flutter-driven FP-OFT known to give better performance~\cite{boudreauParametric2020} are intrinsically not suitable in a horizontal configuration. Indeed, the coupled-flutter instability relies on the coupling of the two degrees of freedom, which depends on a high value of mass and mass imbalance, which also accentuates the impact of buoyancy on the motion of the foil. This ultimately results in a considerably reduced performance. The investigation of a configuration taking into account the findings made in this article remains to be done.

\newpage
\section*{Competing interests statement}
The authors declare there are no competing interests.
 
\section*{Author contribution statement}
A. Roy-Saillant:
data curation,
formal analysis,
investigation,
methodology,
software,
validation,
visualization,
writing -- original draft.
G. Dumas:
conceptualization,
funding acquisition,
project administration,
supervision,
writing -- review \& editing.
L. Duarte:
conceptualization,
supervision.
G. Dellinger:
conceptualization,
funding acquisition,
project administration,
supervision.
M. Olivier:
conceptualization,
formal analysis,
funding acquisition,
investigation,
methodology,
project administration,
resources,
software,
supervision,
writing -- review \& editing.
 
\section*{Funding statement}
This work was supported by the Natural Sciences and Engineering Research Council of Canada (grant No. NSERC-RGPIN-2019-04489), the Fonds de Recherche du Québec -- Nature et technologies (grant No. 301088) and the Digital Research Alliance of Canada. The authors acknowledge the use of Copilot for writing assistance and language improvement.
 
\section*{Data availability statement}
Data generated or analyzed during this study are available from the corresponding author upon reasonable request.

\newpage
\bibliographystyle{CSME_style}
\bibliography{CSME-Paper_ARS}

\end{document}